\documentclass[useAMS]{mn2e}
\usepackage{times,psfig} 

\title[Brief history of the metal accumulation in the ICM]
{Brief history of the metal accumulation in the intracluster medium}
 
\author[] 
{\parbox[]{6.in} {S. Ettori \\ 
\footnotesize 
INAF, Osservatorio Astronomico di Bologna, via Ranzani 1, I-40127 
Bologna, Italy (stefano.ettori@bo.astro.it) }}
 
\date{MNRAS, in press}

\begin{document} 
\maketitle 

\begin{abstract} 
We use models of the rates of Type Ia supernovae (SNe Ia) and core-collapse supernovae, 
built in such a way that both are consistent with recent observational
constraints at $z \la 1.6$ and can reproduce the measured cosmic star formation 
rate, to recover the history of the metals accumulation in the intra-cluster
medium.
We show that these SN rates, in unit of SN number per comoving volume
and rest-frame year, provide on average a total amount of Iron that is 
marginally consistent with the value measured in galaxy clusters in the redshift 
range $0-1$, and a relative evolution with redshift that is in agreement 
with the observational constraints up to $z \approx 1.2$.  
Moreover, we verify that the predicted metals to Iron ratios reproduce
the measurements obtained in nearby clusters through X-ray analysis,
implying that (1) about half of the Iron mass and
$\ga$ 75 per cent of the Nickel mass observed locally are produced by SN Ia ejecta, 
(2) the SN Ia contribution to the metal budget decreases steeply with redshift 
and by $z \approx 1$ is already less than half of the local amount
and (3) a transition in the abundance ratios relative to the Iron is present
between redshifts $\sim 0.5$ and $1.4$, with core-collapse SN products becoming 
dominant at higher redshifts.
\end{abstract} 
 
 
\section{INTRODUCTION} 

The primordial cosmic gas is composed by Hydrogen (about 75 per cent by mass),
Helium ($\sim 24$ per cent), and traces of other light elements, like
Deuterium, Helium-3, Lithium and Berillium. 
When this gas collapses into the dark matter halos typical of galaxy clusters 
($\ga 10^{14} M_{\odot}$), it undergoes shocks and 
adiabatic compression, reaching densities of about $10^{-3}$ particles
cm$^{-3}$ and temperatures of the order of $10^8$ K.
For these physical conditions, the plasma is optically thin and its  
X-ray emission is dominated by thermal bremsstrahlung processes.
Its enrichment by metals (i.e. elements with atomic number larger than 5)
proceeds by the releases to the medium of the products of the star formation 
activities that take place in the member galaxies.  

An effective way to estimate the metal abundance by number relative to Hydrogen
is to measure the equivalent width of the emission lines above the X-ray thermal 
continuum. 
Nowadays, the Iron abundance is routinely determined in nearby systems thanks,
in particular, to the prominence of the K-shell (i.e. the lower level of the electron 
transition is at the principal quantum number $n=1$) Iron line emission at rest-frame 
energies of 6.6-7.0 keV (Fe XXV and Fe XXVI). 
Furthermore, 
the few X-ray galaxy clusters known at $z>1$ have shown well detected Fe line, given
sufficiently long ($\ga 200$ ksec) {\it Chandra} and {\it XMM-Newton} exposures
(Rosati et al. 2004, Hashimoto et al. 2004, Tozzi et al. 2003).
It is more difficult is to assess the abundance of the other prominent metals that 
should appear in an X-ray spectrum at energies (observer rest frame)
between $\sim 0.5$ and $10$ keV, such as Oxygen (O VIII) at (cluster rest frame) 
0.65 keV, Silicon (Si XIV) at 2.0 keV, Sulfur (S XVI) at 2.6 keV, 
and Nickel (Ni) at 7.8 keV.  

A direct application of the detection of emission lines from these highly ionized
elements is the possibility to describe the way the intracluster medium
(ICM) is enriched of metals,  whether by products from explosions
of supernovae (SN) type Ia, mainly rich in Fe and Ni, 
or by outputs of collapsed massive stars (Core Collapsed SN), 
with relative higher contributions of $\alpha-$elements like O, Si, S.
However, the ability to resolve and measure elemental abundances other than Iron,
which was observed in the early days of the X-ray analysis of galaxy clusters
(e.g. Mitchell et al. 1976, Serlemitsos et al. 1977) has required
a significant improvement in the spectral resolution and sufficiently large 
telescope effective areas that have been reached initially with the X-ray satellite
{\it ASCA} (e.g. Mushotzky et al. 1996, Fukazawa et al. 1998, 
Finoguenov, David \& Ponman 2000) and then with the observatories of the
new generation, such as {\it Chandra} (e.g. Ettori et al. 2002, Sanders et al. 2004)
and {\it XMM-Newton} (e.g. B\"ohringer et al. 2001, 
Gastaldello \& Molendi 2002, Finoguenov et al. 2002). 

In the present work, we reverse the problem and infer from observed and modeled 
SN rates the total and relative amount of metals that should be present in the 
ICM both locally and at high redshifts.
Through our phenomenological approach, we adopt the models of SN rates as a function of
redshift that reproduce well both the very recent observational determinations of SN
rates at $z \ga 0.3$ (Dahlen et al. 2004, Cappellaro et al. 2005)
and the measurements of the star formation rate derived from UV-luminosity
densities and IR data sets. We then compare the products of the enrichment 
process to the constraints obtained in the X-ray band for galaxy clusters 
observed up to $z \la 1.2$.
We note that previous work on the production of Iron in galaxy clusters
by, e.g., Matteucci \& Vettolani (1988), Arnaud et al. (1992),
and Renzini et al. (1993) made use of SN rates that were known, with large
uncertainties, only at $z \approx 0$. However, they concluded
that either the SN Ia rates were higher by a factor of 10 in the past or
the production of Iron in clusters by core-collapsed supernovae (SNe CC) 
would require a very flat Initial Mass Function (IMF). 
An alternative scenario is that SN CC produce 1/4 of the Iron and 
the remnants 3/4 is released from SN Ia, which would also 
be in accordance with the standard chemical model for the galactic halo and
disk, allowing an IMF in elliptical galaxies similar to the one in the
solar neighborhood.

The main purpose of this work is 
to infer the mass of the most relevant and X-ray detectable elements
present in the ICM and their relative abundance as function of redshift
by using reliable (i.e. in agreement with several different observational
constraints) models of SN explosion rates that include a dependence on
the cosmological time. The fact that these rates
are in unit of number per comoving volume and rest-frame time allows us to
use them also over cosmological distances typical of high-redshift systems.
The method presented here is a first attempt to provide well-motivated
and analytic predictions of the expected metal abundances at $z \ga 0.5$
and of the relative role played by SNe Ia and SNe CC in polluting the ICM 
over time.

\begin{table}
\caption{Adopted values for the average atomic weight ($W$), solar abundance by number
with respect to Hydrogen ($A$, from Anders \& Grevesse 1989), total synthesized
isotopic mass per SN event ($m_{Ia}$ and $m_{CC}$ from Table~1 in Nomoto et al. 1997;
nucleosynthesis products of SN Ia are those from the deflagration model W7,
whereas SN CC yields
are integrated over the mass range $10-50 M_{\odot}$ with a Salpeter IMF)
and corresponding abundance ratios by numbers with respect to the Iron,
with each abundance normalized to the solar photospheric value $A$
($Y_i = m_i / m_{\rm Fe} \times W_{\rm Fe}/W_i \times A_{\rm Fe}/A_i$).
}
\begin{tabular}{c c c c c c c}
\hline \\
{\it metal} & $W$ & $A$ & $m_{Ia}$ & $Y_{Ia}$ & $m_{CC}$ & $Y_{CC}$  \\
 & & & $M_{\odot}$ & & $M_{\odot}$ \\
Fe & 55.845 & 4.68e-5 & 0.743 &  $-$ & 0.091 &  $-$  \\
O  & 15.999 & 8.51e-4 & 0.143 & 0.037 & 1.805 & 3.818  \\
Si & 28.086 & 3.55e-5 & 0.153 & 0.538 & 0.122 & 3.526  \\
S  & 32.065 & 1.62e-5 & 0.086 & 0.585 & 0.041 & 2.284  \\
Ni & 58.693 & 1.78e-6 & 0.141 & 4.758 & 0.006 & 1.647  \\
\hline \\
\end{tabular}
\end{table}

\section{Observational constraints on the metal content of galaxy clusters}

The observed Iron mass is obtained as
\begin{equation}
M_{\rm Fe, obs} = 4 \pi A_{\rm Fe} W_{\rm Fe} \int_0^R Z_{\rm Fe}(r) \rho_{\rm H}(r) r^2 dr,
\end{equation}
where $Z_{\rm Fe}(r)$ is the Iron abundance relative to the solar value $A_{\rm Fe}$
measured in the ICM as function of radius, 
$\rho_{\rm H}(r) = \rho_{\rm gas}(r) / (2.2 \mu)$ is the Hydrogen density, $\mu$ is the mean
molecular weight and has a value of 0.60 for a typical fully ionized cluster plasma
with a number density $n_{\rm gas} = n_{\rm H} +n_{\rm e} = 2.2 n_{\rm H}$.
Our estimates refer to a cosmology with parameters 
$(H_0, \Omega_{\rm m}, \Omega_{\Lambda})$ equal to $(70, 0.3, 0.7)$.
 
\begin{figure}
\psfig{figure=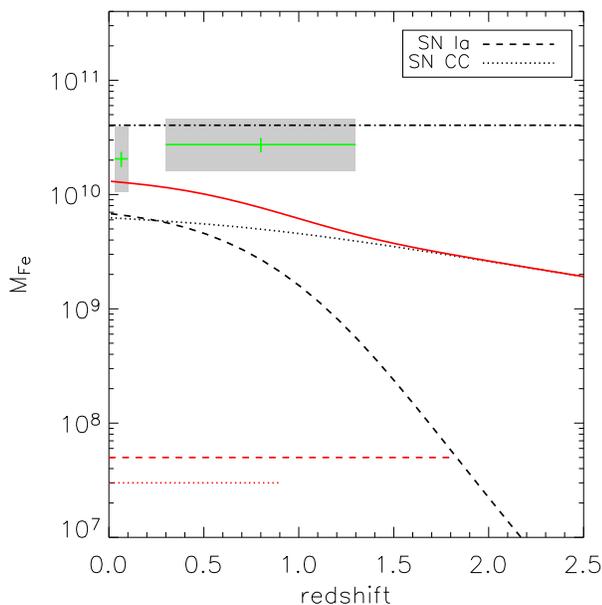,width=0.5\textwidth}
\caption[]{Accumulation history of the Iron mass from $z_F = 10$ to $z=0$, as 
obtained from equations~(2) and (3). The solid line shows the sum of SN Ia
({\it dashed line}) and SN CC ({\it dotted line}) contribution.
Horizontal lines indicate the redshift region where the SN rates 
are actually measured (Dahlen et al. 2004).
The accumulation history is here compared to the total Iron mass measured
in samples of local (De Grandi et al. 2004) and high$-z$ (Tozzi et al. 2003)
galaxy clusters for a typical object at 5 keV.
Error bars on the mean observed value and scatter ({\it shaded region})
around it are shown (see text in Section~2). 
The {\it dot--dashed line} indicates the expected Iron mass for a typical
5 keV cluster with $M_{\rm vir} = 10^{15} M_{\odot}$, a gas mass fraction 
of 10 per cent and Iron abundance of $0.3 A_{\odot}$.  
} \label{fig:mfe_z} \end{figure}

De Grandi et al. (2004) have correlated the total amount of $M_{\rm Fe}$ 
observed with the X-ray satellite {\it BeppoSAX} in 22 nearby hot ($kT > 2$
keV) galaxy clusters with the gas temperature. 
For the assumed cosmology, we obtain a best-fitting result of
$\log(M_{\rm Fe}/10^{10} M_{\odot})
= 0.33 (\pm 0.07) +1.62 (\pm 0.53) \times \log(kT/5 {\rm keV})$ (scatter of 0.28).
This fit considers quantities estimated at the radius $R_{500}$, where the
cluster overdensity is 500 times the critical density for an Einstein-de Sitter
universe (i.e. an overdensity of 285 at $z=0$ for the cosmology assumed here)
and corresponds to about $0.6 \times R_{\rm vir}$.
The overall results presented in this paper do not depend upon the cluster
gas temperature, although a value of $5$ keV is hereafter adopted as reference.
Therefore, an Iron mass of $(2.1 \pm 0.3) \times 10^{10} M_{\odot}$
is associated with a $5$-keV cluster within $R_{500}$ at a median redshift of 
$0.05$, with a scatter in the range of $(1.1, 3.9) \times 10^{10} M_{\odot}$.
At redshift between $0.3$ and $1.3$, for a sample of 19 objects observed 
with {\it Chandra} and presented in Tozzi et al. (2003) and Ettori et al. (2004a),
that have a detection significant at the level of $2 \sigma$ of the Iron line emission 
and for which a single emission-weighted temperature and  
the radial profile of the gas density were measured, 
we obtain a best-fitting
$\log(M_{\rm Fe}/10^{10} M_{\odot}) = 0.66 (\pm 0.06) +1.47 (\pm 0.48) 
\times \log(kT/5 {\rm keV})$ (scatter of 0.23) and infer a total 
Iron mass for a $5$ keV cluster at the median $z=0.63$ 
of $M_{\rm Fe} (<R_{500}) = (2.7 \pm 0.4) \times 10^{10} M_{\odot}$,
with a scatter limited within the values of $(1.6, 4.6) \times 10^{10} M_{\odot}$. 
These values are plotted in Fig.~\ref{fig:mfe_z}.
We assume that most (if not all) of the mass of the metals is 
located within $R_{500}$, even though this radius encloses just about 13 per cent
of the virial volume ($R_{500} \approx 0.5 R_{\rm vir}$).
A further contribution from the outer cluster regions is however marginal:
from numerical simulation (e.g. Ettori et al. 2004b) and extrapolated data
(e.g. Arnaud, Pointecouteau \& Pratt 2005), one can estimate
that $M_{500} / M_{\rm vir} \approx 0.6$ and, by assuming that in the
outskirts (1) the metallicity is $<0.1$ times the solar value, $A_{\odot}$, 
whereas its mean value in the central regions is about $0.3 A_{\odot}$,
and (2) the gas follows the dark matter distribution, one can evaluate
that $M_{\rm Fe} (R_{500} < r < R_{\rm vir}) / M_{\rm Fe} (r < R_{500}) 
\la 0.2$.

Apart from Fe, 
the most prominent lines detectable in X-ray spectra, like O, Si, S and Ni, 
were investigated initially with data from {\it ASCA} (an analysis of the 
average abundance of all the galaxy clusters in the {\it ASCA} archive is
presented in Baumgartner et al. 2005) and more recently with the larger 
effective area and improved spectral and spatial capabilities of {\it XMM-Newton} 
(e.g. Tamura et al. 2004 present statistics of abundance ratios, also resolved
spatially, for a sample of 19 nearby galaxy clusters). 
In Fig.~\ref{fig:yfe_z0}, we summarize these results 
by plotting the relative number abundance ratios. 
The Iron abundance is about $0.3 A_{\odot}$, 
although the {\it ASCA} measurements have a very narrow distribution around 
the mean of $0.27$ ($\pm 0.01$ at 90 per cent level of confidence) that 
departs significantly from the spatially resolved {\it XMM-Newton} estimates 
(only in the outer radius, between 200 and 500 $h^{-1}$ kpc, where
the Fe/H decreases by about a factor of 2 with respect to the central estimate, 
it becomes consistent with the {\it ASCA} mean with a value of $0.32 \pm 0.08$).
Moreover, the O/Fe ratio is about solar, whereas the Si/Fe and S/Fe
ratios are roughly super-solar and sub-solar, respectively.
The average Ni/Fe ratio in the {\it ASCA} sample is about $3.4 A_{\odot}$, 
well in agreement with the original {\it ASCA} determination in the data 
of the Perseus cluster by Dupke \& Arnaud (2001) and consistent with the
{\it XMM-Newton} constraints in Gastaldello \& Molendi (2004) and with
the {\it BeppoSAX} results for a sample of 22 objects 
(Fig.~6 in De Grandi \& Molendi 2002). On the other hand, it is worth
mentioning that the measurements of the Nickel abundance becomes
reliable only in hot ($kT \ga 4$ keV) systems where the excitation 
of (mainly) K-shell lines makes them detectable in the spectrum around 8 keV,
in a region less contaminated from blends with Iron but highly affected 
from a proper background subtraction.
These difficulties put the actual limitations on solid Ni measurements
for a large data set.

Given the differences between the two data sets, and
the fact that the {\it XMM-Newton} estimates are spatially resolved and thus sensitive
to the presence of any gradient that instead is washed out in the global
measurement shown here from {\it ASCA}, 
they are considered independently in the following analysis.

\section{Accumulating the metals}

\begin{figure}
\psfig{figure=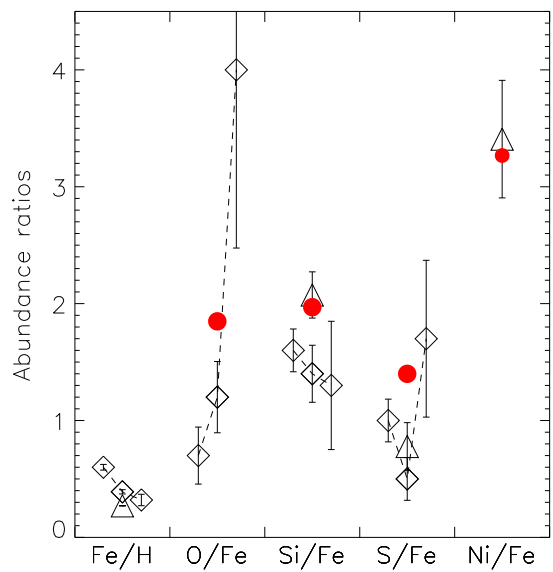,width=0.5\textwidth}
\caption[]{Predictions ({\it big dots}) of the local metal yields 
compared with the X-ray measurements in Tamura et al. (2004; case for a {\it medium-T} 
cluster in Table~6 with estimates in the three radial bins $0-50$, $50-200$ and 
$200-500 h^{-1}$ kpc; {\it diamonds}) 
and Baumgartner et al. (2005; 4.5 keV bin in Table~4; {\it triangles}). 
Error bars at $1 \sigma$ level are indicated.
The abundance measurements are relative to the Anders \&
Grevesse (1989) solar photospheric values quoted in Table~1.
} \label{fig:yfe_z0} \end{figure}

\begin{figure}
\psfig{figure=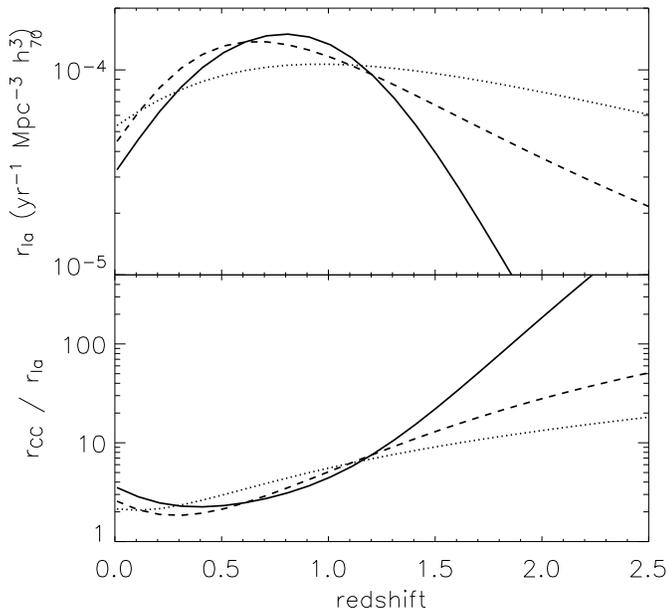,width=0.5\textwidth}
\caption[]{SN Ia rates and number ratio of SNe CC to Ia as a function 
of redshift for the delay time distribution functions considered in the 
present work.
({\it Solid line}) A ``narrow" Gaussian with $\tau = 4$ Gyr and
$\sigma_{t_d} = 0.2 \tau$, which is the one adopted here;
({\it dashed line}) a ``wide" Gaussian with $\tau = 4$ Gyr and
$\sigma_{t_d} = 0.5 \tau$; ({\it dotted line}) a exponential
form, $\phi(t_d) \propto \exp(-t_d / \tau)$, with $\tau = 5$ Gyr.
A similar figure is presented and discussed in Dahlen et al. (2004).
} \label{fig:r_cc_ia} \end{figure}

\begin{figure}
\psfig{figure=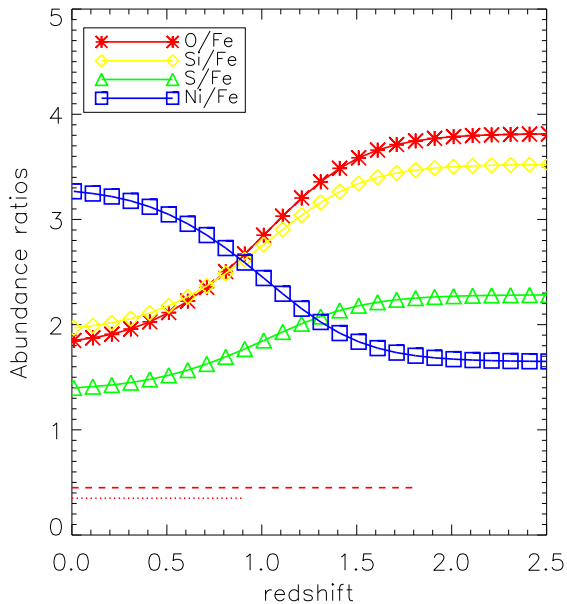,width=0.5\textwidth}
\caption[]{Metal abundance ratios as a function of redshift.
Horizontal lines indicate the redshift region where the SN rates
are actually measured (Dahlen et al. 2004).
} \label{fig:yfe_z} \end{figure}

The total predicted Iron mass is obtained as
\begin{eqnarray}
M_{\rm Fe, SN} & = & M_{\rm Fe, Ia} +M_{\rm Fe, CC}  \nonumber \\
 & = & \sum_{dt, dV} m_{\rm Fe, Ia} \times r_{\rm Ia}(dt) \times dt \times dV \nonumber \\  
& & +\sum_{dt, dV} m_{\rm Fe, CC} \times r_{\rm CC}(dt) \times dt \times dV  
\end{eqnarray}
where we assume $m_{\rm Fe, Ia}$ and $m_{\rm Fe, CC}$ as quoted in Table~1
(from Nomoto et al. 1997) and an interval time at given redshift
equals to $dt(z0,z1) = H_0^{-1} \int_{z0}^{z1} (1+z)^{-1}  
\left[\Omega_{\rm m} (1+z)^3 +(1-\Omega_{\rm m}-\Omega_{\Lambda}) (1+z)^2
+\Omega_{\Lambda} \right]^{-0.5} dz$.
As cluster volume, we define the volume corresponding
to the spherical region that encompasses the cosmic background density,
$\rho_{\rm b} = 3 H_0^2 / (8 \pi G) \times \Omega_{\rm m}$ 
($\approx 4 \times 10^{10} M_{\odot}$ Mpc$^{-3}$ for the assumed cosmology),
with a cumulative mass of $M_{\rm vir} \approx 6.8 
(kT / 5{\rm keV})^{3/2} \times 10^{14} M_{\odot}$, 
as inferred from the best-fit results on the observed
properties of X-ray clusters with $kT > 3.5$ keV in Arnaud et al. (2005)
and adopting $M_{500}/ M_{\rm vir} \approx 0.6$: 
$dV = M_{\rm vir} / \rho_{\rm b} = 24,514 \ \frac{M_{\rm vir}}{10^{15} M_{\odot}} 
\left( \frac{kT}{5{\rm keV}} \right)^{3/2} \ h_{70}^{-3}$ Mpc$^3$.
It is worth noticing that the $M_{\rm vir}-T$ relation for simulated 
galaxy clusters (e.g. Ettori et al. 2004b), with an emission-weighted 
temperature greater than $3.5$ keV, provides a typical 
virial mass (and corresponding volume) that is larger than the above value
by about 20 per cent.
We discuss in Section~4 the effect of this discrepancy on the overall results.
The volume measured at $z=0$ is considered as the dimension of the region 
involved in the formation of a typical galaxy cluster and is
maintained constant in redshift. 

The supernovae rates, i.e. {\it the number of SN per rest-frame time 
and comoving volume}, come from models that fit well the observed values
and reproduce properly the Star Formation Rate (SFR) as widely discussed in
Strolger et al. (2004) and Dahlen et al. (2004).
In summary, once a functional form is defined to describe the evolution
with time of the SFR (see equation 5 in Strolger et al. 2004 with parameters
for the extinction-corrected model, called M1; the uncorrected M2 model just 
reduces the total amount of metals released by a factor of $\sim 4$), the
corresponding SN rates can be defined as
\begin{eqnarray}
r_{\rm Ia}(t) & = & \nu \int_{t_F}^t SFR(t') \times \phi(t-t') \ dt'  \nonumber \\
r_{\rm CC}(t) & = & k \times h^2 \times SFR(t)
\end{eqnarray}
where the normalization $\nu$ indicates how many SNe Ia explode per unit
formed stellar mass and is function of the delay time distribution function
$\phi(t_d)$ that represents the relative number exploded at a time $t_d$
since a single burst of star formation; $t_F$ is the time of formation of the
first stars and corresponds to $z_F = 10$; the normalization $k$ is the number    
of CC progenitors per unit of formed stellar mass; 
$h$ is the Hubble constant in unit of 100 km s$^{-1}$ Mpc$^{-1}$.
In the following analysis, we adopt a ``narrow" Gaussian form for $\phi(t_d)$,
with a delay time $\tau = 4$ Gyr and $\sigma_{t_d} = 0.2 \tau$ that better
fit both the local ($z \la 0.1$) rates and the GOODS data (Strongler et al. 2004,
Dahlen et al. 2004) and fix $\nu = 0.0010$. 
We assume $k = 0.0069 M_{\odot}^{-1}$ as appropriated 
for CC progenitors masses in the range $8 M_{\odot} \la M \la 50 M_{\odot}$
with a Salpeter (1955) Initial Mass Function (IMF).
We show in Fig.~\ref{fig:r_cc_ia} the relative rates of SNe CC and Ia as
function of redshift: this ratio ranges between 2 and 3 up to
$z \approx 1$ and increases steeply at higher redshifts, with the ``narrow" 
Gaussian function that provides a SN Ia rate at $z \approx 2$ lower by an 
order of magnitude than the other delay time distribution functions.

Regarding the adopted delay time, we mention here that Gal-Yam \& Maoz (2004) and 
Maoz \& Gal-Yam (2004) support the argument that some current
data on the cosmic (field) star formation history require for the observed
rate of SNe Ia a delay time larger than 3 Gyr. Combining this with the low
observed rate of cluster SN Ia at $z<1$, they conclude that
the measured amount of Iron in clusters cannot be produced mainly by
SNe Ia explosions but, more probably, must be produced via outputs of SNe CC 
originating from a top-heavy IMF. 
The reader is refereed to the above-mentioned work for a 
detailed discussion on the effects of the assumed IMF on the SN rates.
Our approach here is to use the best phenomenological models of 
cosmic star formation history to infer the corresponding cluster
metal accumulation history and to compare some predictions with 
the observational constraints available. 

To produce the cluster metal accumulation history, we have to take
into account the fraction of the total produced Iron mass that 
remains locked up in the stars. 
The fraction of the total Iron that is then released to the ICM is estimated 
by considering that $M_{\rm Fe, tot} \approx Z_{\rm ICM} M_{\rm ICM} +
Z_{\rm star} M_{\rm star}$.
We adopt $Z_{\rm ICM} \approx 0.3 A_{\odot}$ (see Fig.~\ref{fig:yfe_z0}), 
$M_{\rm star} \approx 0.13 M_{\rm ICM}$ (from equations 1 and 10 in 
Lin, Mohr \& Stanford 2003 for a 5 keV cluster) and 
$Z_{\rm star} = 1.38 A_{\odot}$ (from interpolation of the results in 
Lin, Mohr \& Stanford 2003; see also Renzini 2003, Portinari et al. 2004).
Therefore, $M_{\rm Fe, ICM} = M_{\rm Fe, tot} / (1 + f_Z)$, where $f_Z =
M_{\rm star} / M_{\rm ICM} \times Z_{\rm star} / Z_{\rm ICM} \approx 0.59$.
Considering its weak dependence upon time (see, e.g., Table~2 and Fig.~3 in 
Portinari et al. 2004), we neglect the evolution of the locked-up
fraction.

In Fig.~\ref{fig:mfe_z}, we plot the results of the expected Iron mass
compared to the observational results discussed in Section~2. 
The predicted amount is well in agreement with the observed Iron mass 
in local clusters. At a median redshift of 0.05, the De Grandi et al. sample
has a mean $M_{\rm Fe}$ associated to a 5 keV object that is between 30 
(when the $M_{\rm vir}-T$ relation from simulations is adopted) and 60
(with the observed $M_{\rm vir}-T$ relation) per cent
higher than the Iron mass accumulated in the ICM 
through SN activities from $z=10$.
The latter value is, however, well within the scatter in the observed distribution.
At higher redshift, the observed central value in the Tozzi et al. sample
of Fe abundance measurements is a factor of about 3 higher than the predicted one.

It is worth noticing that this result does not depend significantly upon either
the delay time distribution function adopted to calculate $r_{\rm Ia}$ or the
SN CC and Ia compilations assumed. The changes in the total Iron mass accumulated
at $z=0.05$ are in the order of 10 per cent when different $\phi(t_d)$ 
are considered and of about 40 per cent (i.e. the ratio of the observed and
the predicted values ranges between 1.0--1.2 and 1.4--1.8; see Table~\ref{tab:test}) 
when we use the calculations of SN CC explosions in two extreme cases presented 
in Woosley \& Weaver (1995) and described at the end of this Section.

Note that, if the SN Ia rate in number per comoving volume and rest-frame year is 
converted to supernova unit (1 SNu = 1 SN per 100 year per $10^{10} L_{\odot, B}$)
with an assumed local B-band luminosity density of $2 \times 10^8 h L_{\odot}$
Mpc$^{-3}$ which evolves as $(1+z)^{1.9}$ (see Dahlen et al. 2004 for the caveats 
in using such conversion for rates estimated on cosmological distances), 
we obtain a local ($z \approx 0.05$) rate of $0.25$ and $0.79$ SNu for 
SN Ia and CC, respectively. By integrating the SNu values over the cosmic time,  
we measure a local $M_{\rm Fe} / L_B$ of $\sim 0.0020$ and $0.0010 h_{70}^2$
as due to SN Ia and CC, respectively, with a dependence upon the redshift very 
similar to what is shown in Fig.~\ref{fig:mfe_z}. 
The sum of these values is still a factor between 3 and 5 lower than 
the present estimate in galaxy clusters: this estimate, however, is affected from 
the extension of the cluster region considered to measure B-band luminosities 
and Iron mass distribution (see discussion on the uncertainties of these measurements
in De Grandi et al. 2004, Sect.3.2). Therefore, even though we solve most 
of the discrepancy between predicted and measured Iron mass in typical 
galaxy clusters through models of SN rates that make use of the number
of SN per comoving volume and rest-frame time, some difficulties remain
in recovering the observed $M_{\rm Fe} / L_B$ ratio through
a direct, but not straightforward (given the evolution with redshift
of the B-band luminosity density), conversion to the number of SN per unit 
of B-band luminosity (e.g. Arnaud et al. 1992, Renzini et al. 1993). 
Regarding the latter issue, 
note that the suggested solution of an increase of the SN Ia rate in E/S0 galaxies
by a factor of 5--10 at higher$-z$ is partially taken into account 
in the models presented here (see, e.g., Fig.~\ref{fig:r_cc_ia}), where
the SN Ia rate is enhanced by a factor of $\sim 5$ from $z=0$ to $z \approx 0.8$, 
but then decreases rapidly beyond it.
 
We can now extend these considerations to the enrichment history of the
metals other than Fe. To infer these, we adopt, as done in the X-ray analyses
presented above, the solar abundances tabulated in Anders \& Grevesse 
(1989) and the most recent and widely used theoretical metal yields of 
Nomoto et al. (1997) for 
(i) SNe CC integrated between $10$ and $50 M_{\odot}$ for a Salpeter IMF 
(see Nakamura et al. 1999 on the uncertainties related 
to the adopted mass cut that can change the Nickel yield by a factor of 2)
and (ii) SNe Ia exploding with fast deflagration according to the W7 model.
All the numerical values considered here are presented in Table~1.
We note that the values presented in Nomoto et al. (1997) are plotted
relative to the solar photospheric abundance in Anders \& Grevesse (1989).
The latter values have been revised to match the meteoritic determinations, 
as summarized in Grevesse \& Sauval (1998), and require 
conversion factors of $(0.676, 0.794, 1, 1.321, 1)$ to correct
the original photospheric abundance for Fe, O, Si, S and Ni, respectively.
We have, however, adopted the Anders \& Grevesse values for consistency
between models and observational constraints. 

Given a metal $i$, its total mass is estimated as
\begin{eqnarray}
M_i  & = & M_{i, Ia} + M_{i, CC} \nonumber \\
 & = & \sum_{dt, dV} m_{\rm i, Ia} \times r_{\rm Ia}(dt) \times dt \times dV \nonumber\\  
& & +\sum_{dt, dV} m_{\rm i, CC} \times r_{\rm CC}(dt) \times dt \times dV \nonumber\\   
 & = & \left(M_{{\rm Fe}, Ia} \ Y_{Ia} + M_{{\rm Fe}, CC} \ Y_{CC}
\right) \frac{W_i \ A_i}{W_{\rm Fe} \ A_{\rm Fe}}, 
\end{eqnarray} 
where the synthesized masses for each element of interest, $m_i$, or the
corresponding abundance yields, $Y$, are presented in Table~1. 

One direct way to check the consistency of these models is to compare
the ratio between some of the most prominent elements detectable in X-ray spectra  
and the Iron. To this purpose, we overplot in Fig.~\ref{fig:yfe_z0}
the abundance ratios inferred from the models to the observational constraints
on the O/Fe, Si/Fe, S/Fe and Ni/Fe discussed in Section~2.
The agreement is remarkably good for what concerns the {\it XMM-Newton} means and
the {\it ASCA} Ni/Fe measurement. A larger departure appears between the model
prediction and the result for the S/Fe estimate from {\it ASCA}.
We can evaluate the agreement by calculating the $\chi^2$ given the observational
constraints (with the relative errors) and the predicted values.
We obtain a $\chi^2 = 3.1$ (3 degrees of freedom) for the {\it XMM-Newton} data, 
which is statistically acceptable  [$P(\chi^2 > \chi^2_{\rm obs}) = 0.63$] and is,
however, the largest value measured for any delay time distribution functions:
the ``wide" Gaussian and the $e-$folding form (see caption of 
Fig.~\ref{fig:r_cc_ia}) give a $\chi^2$ of $2.2$ and $2.3$, respectively
(see Table~\ref{tab:test}).  
It is worth noting that the Si/Fe ratio alone contributes about 
$2.2$ to the total $\chi^2$, owing to the over-predicted amount of Silicon 
by $\sim 40$ per cent.
On the other hand, the {\it ASCA} measurements provide a $\chi^2$ of about
10.0 [3 degrees of freedom, $P(\chi^2 > \chi^2_{\rm obs}) = 0.98$], mainly
due to the over-prediction of the S/Fe ratio by 80 per cent.
We have also considered alternative SN CC and Ia compilations 
to the Nomoto et al. one, adopted here as reference.
Following the discussion in Gibson, Loewenstein \& Mushotzky (1997) on the 
uncertainties related to the physics of exploding massive stars, 
we consider two extreme conditions for the explosion of SN CC
as indicated in Woosley and Weaver (1995), namely case A with metallicity 
equals to $10^{-4}$ solar and case B with solar metallicity.
Both the models provide best-fit results worse than the Nomoto et al. one,
with a $\chi^2$ of 9.3 and 3.1 (case A and B, respectively) when
compared to the Tamura et al. abundance ratios and 42.5 and 21.0 when 
compared to the Baumgartner et al. values (see Table~\ref{tab:test}).
Moreover, we consider SN Ia models with a delayed detonation induced
from deflagration at low density layers (model WDD2 in Nomoto et al. 1997)
that seem to reproduce well the observed yields in the outer regions
of M87 (Finoguenov et al. 2002). 
This model doubles the amount of Si ($m_{\rm Si} =
0.27 M_{\odot}, Y = 1.01$) and S ($m_{\rm S} = 0.17 M_{\odot}, Y = 1.20$)
and reduces the Ni ejecta by 70 per cent 
($m_{\rm Ni} = 0.04 M_{\odot}, Y = 1.40$)
with respect to W7, amplifying any mismatch with the observational 
mean values so that the $\chi^2$ increases to $\sim$ 8 and 38 when model
predictions are compared with data from Tamura et al. and Baumgartner et al., 
respectively. We thus exclude the possibility that 
this model of SN Ia explosions can explain the present results on the overall
enrichment history of galaxy clusters better than W7.

\begin{table*} 
\caption{Tests performed on the different SN Ia (W7, WDD2) and 
CC (N97, WW95A, WW95B) models and delay time distribution function 
$\phi(t_d)$ (narrow Gaussian, wide Gaussian, exponential form) adopted. 
The first column indicates the change with respect to the favored model 
shown in the first row:
W7 deflagration model of SN Ia products, Nomoto et al. 1997 (N97) compilation
of SN CC outputs and a ``narrow'' Gaussian delay time distribution function.
The 5 tests done are: comparison between the observational constraints 
(from De Grandi et al. 2004 and Tozzi et al. 2003) on the total amount of Iron 
mass, $\hat{M}_{\rm Fe}$, produced at the median redshift of $0.05$ and $0.63$ 
in a typical cluster of 5 keV and the predicted value, $M_{\rm Fe}$, both
using a $M_{\rm vir}-T$ relation from simulations (low end) and
from observations (high end; 2nd and 3rd column); comparison of
the relative change of the Iron abundance with redshift with 49 measurements
done between redshift $0.3$ and $1.3$ (Fig.~\ref{fig:fe_z}; 4th column);
comparison of the abundance ratios observed in a sample of nearby clusters 
with {\it XMM-Newton} (O/Fe, Si/Fe, S/Fe; Tamura et al. 2004; 5th column) 
and {\it ASCA} (Si/Fe, S/Fe, Ni/Fe; Baumgartner et al. 2005; 6th column).
}
\begin{tabular}{l c c c c c c c}
\hline \\ 
models  & & $\frac{\hat{M}_{\rm Fe}}{M_{\rm Fe}}$ & $\frac{\hat{M}_{\rm Fe}}{M_{\rm Fe}}$ & 
 & $\chi^2$ (dof) & $\chi^2$ (dof) & $\chi^2$ (dof) \\
(Ia, CC, $\phi(t_d)$) & & $z=0.05$ & $z=0.63$ & & vs $Fe(z)$ & vs {\it XMM} & vs {\it ASCA}  \\ 
 & \\ 
\hline \\
(W7, N97, narrow) & & 1.29--1.58 & 2.58--3.16 & & 0.64 (5) & 3.1 (3) & 10.0 (3)  \\
 & & \\ 
(WDD2, ..., ...)  & & 1.33--1.62 & 2.64--3.22 & & 0.64 (5) & 8.1 (3) & 37.5 (3)  \\
(..., WW95A, ...)  & & 1.43--1.75 & 2.94--3.59 & & 0.66 (5) & 9.3 (3) & 42.5 (3) \\
(..., WW95B, ...)  & & 1.02--1.24 & 1.92--2.35 & & 0.66 (5) & 3.1 (3) & 21.0 (3) \\
(..., ..., wide) & &  1.20--1.47 & 2.48--3.03 & & 0.68 (5) & 2.2 (3) & 8.9 (3)  \\
(..., ..., $e-$fol)  & & 1.22--1.49 & 2.33--2.85 & & 0.74 (5) & 2.3 (3) & 9.0 (3) \\
\hline \\ 
\end{tabular}
\label{tab:test}
\end{table*}

\begin{figure}
\vbox{
 \psfig{figure=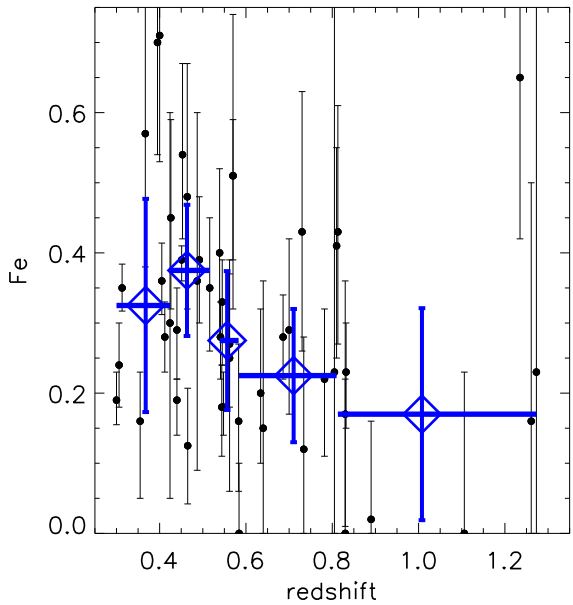,width=0.5\textwidth}
 \psfig{figure=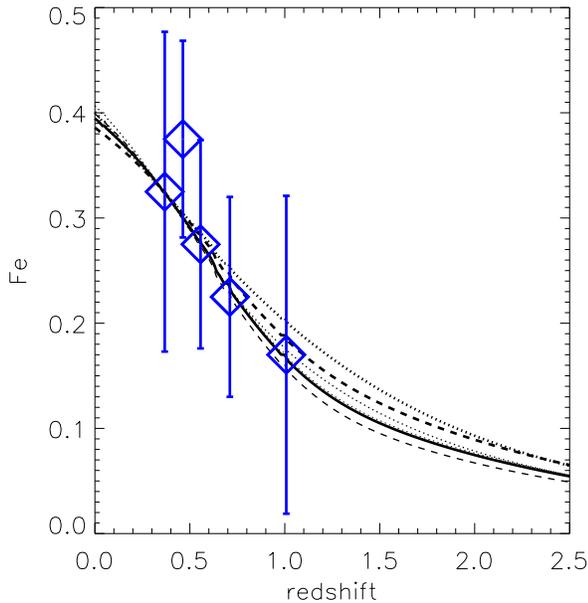,width=0.5\textwidth}
} \caption[]{{\it (Upper panel)} Distribution of the metal abundance
as a function of redshift for a sample of {\it Chandra} exposures
of 49 galaxy clusters at $z>0.3$ and gas temperatures between
3.4 and 15.5 keV (Balestra, Tozzi, Rosati, Ettori
et al. in preparation, which will present the metallicity measurements
for a sample larger by a factor of $\sim 3$ than the one presented in
Tozzi et al. 2003, with upgraded calibration files and extended statistical
analysis). 
The {\it diamonds} refer to the median distribution of 10 Iron
abundance measurements per redshift bin.
{\it (Lower panel)} Predicted and observed Iron abundance as a function
of redshift. The {\it diamonds} refer to the median distribution shown
in the upper panel.
The normalization of the predicted evolution is here fixed equal to the
first measured bin at $z\approx0.4$.
The lines refer to different SN explosion models and delay
time distribution functions with respect to the reference values indicated by the
{\it solid} line ({\it dashed line}: Woosley \& Weaver SN CC models,
case A -thin line- and B -thick line-;
{\it dotted line}, from the thinnest to the thickest line: WDD2 SN Ia model,
wide Gaussian and $e-$folding form).
} \label{fig:fe_z} \end{figure}

\begin{figure}
\psfig{figure=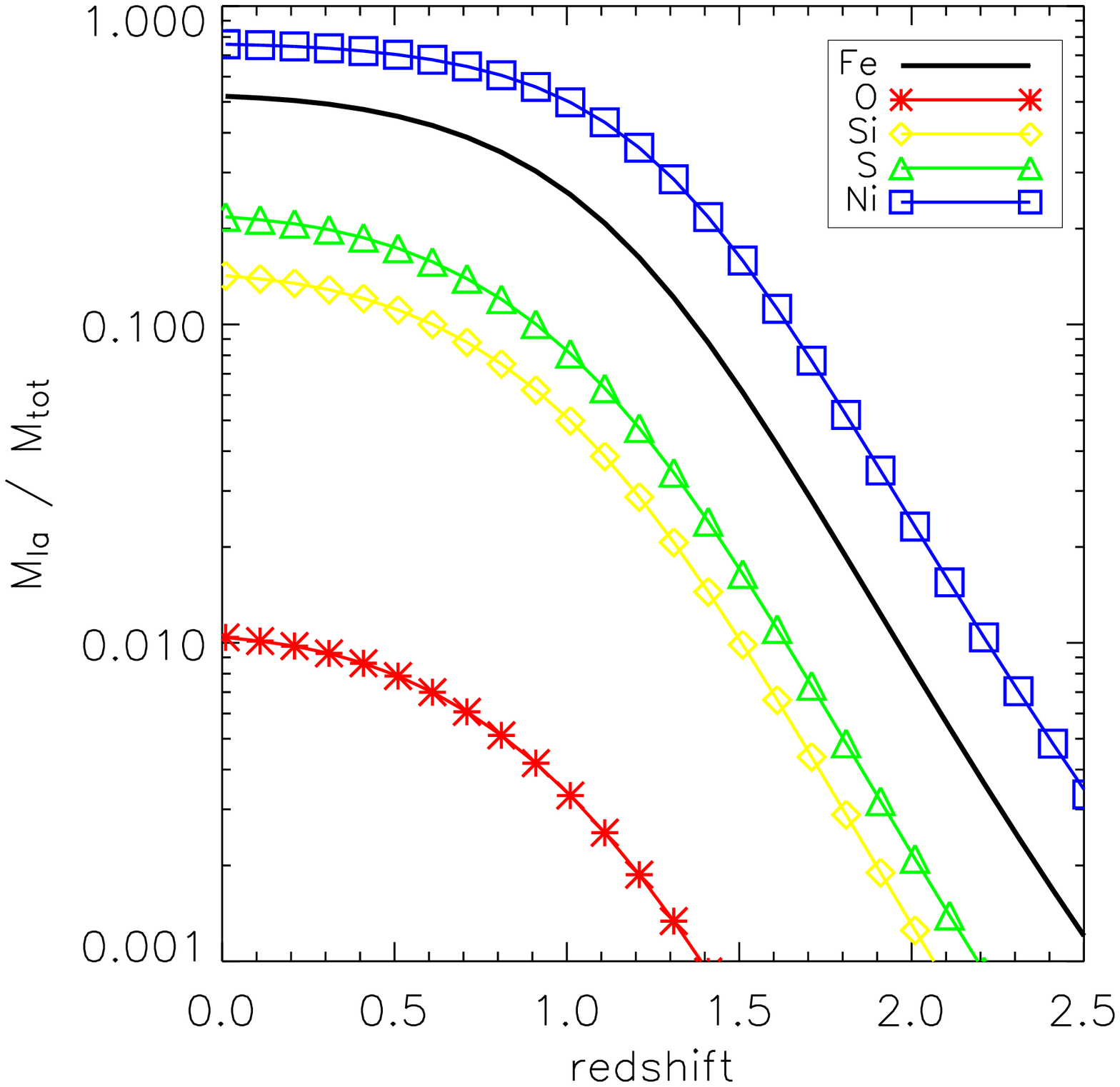,width=0.5\textwidth}
\caption[]{Metal mass fraction resulting from SN Ia ejecta as a function of redshift.
} \label{fig:mia_z} \end{figure}

We can now trace back the abundance ratios as function of redshift
(see Fig.~\ref{fig:yfe_z} and \ref{fig:fe_z}) and, given this history of the
metals accumulation in galaxy clusters, plot
in Fig.~\ref{fig:mia_z} the relative contribution in mass of the SN Ia products
to the total material of each elements investigated (i.e. O, Si, S, Fe and Ni)
available in the cluster baryon budget as function of redshift.

Locally, 51 per cent of Iron, 1 per cent of Oxygen, 14 per cent of Silicon, 
21 per cent of Sulfur and 75 per cent of Nichel are produced from SN Ia
explosions. These values change only by few ($\la 3$) per cent by adopting 
different delay time distribution functions and by $\la 10$ per cent when
the two other extreme cases of SN CC explosion are considered.
We conclude that these predictions for
nearby clusters are particularly stable and robust. 
As expected from the relative increase of SNe CC, the SN Ia contribution to 
the metal budget decreases significantly at higher redshifts,
becoming responsible for less than half of the metal masses
observed locally at $z \approx 1$.
More interestingly, the predicted evolution in the Iron abundance 
(Fig.~\ref{fig:fe_z}), although significantly steepens up to $z \approx 1$ 
where the expected value is half of the local one, is still consistent 
with observational constraints obtained from a sample of {\it Chandra} 
exposures of 49 clusters at redshift $>0.3$ (Fig.~\ref{fig:fe_z}).

\section{Summary and Discussion}

We summarize here our main findings on the history of the metals 
accumulation in the ICM.
By using the rates of SNe Ia and CC as observed (at $z<1.6$ and $z<0.7$,
respectively) and modelled from the cosmic star formation rate derived
from UV-luminosity densities and IR data sets (Dahlen et al. 2004, 
Strolger et al. 2004) and adopting theoretical yields 
(as described in Table~1), we infer how the metals masses in the ICM
are expected to accumulate as a function of the redshift.

We find that these models predict that the total Iron mass accumulated 
in massive galaxy clusters through SN activities is between 24 and 75 
per cent (235--359 per cent) lower than the Iron mass estimated 
in local (high$-z$) systems as determined through the equivalent width
of the emission lines detected in X-ray spectra
(e.g. De Grandi et al. 2004, Tozzi et al. 2003).
This discrepancy is reduced by about 20 per cent when the 
cluster volume used to accumulate the SN products as a function 
of time is defined by adopting an average virial mass 
from numerical simulations instead of the value extrapolated 
from the observed X-ray scaling relationships
(e.g. Ettori et al. 2004b, Arnaud et al. 2005).
Because of this kind of uncertainty in relating the cluster gas 
temperature to the associated virial and Iron masses, we conclude
that the agreement between the expected and measured Iron mass, even
though marginal, is acceptable within the observed scatter. 
Moreover, we can reproduce the relative number abundance
of the most prominent metals detectable in an X-ray spectrum, 
such as Oxygen, Silicon, Sulfur and Nickel with respect to the
Iron as estimated 
for nearby bright objects observed both with {\it ASCA} 
(Baumgartner et al. 2005) and {\it XMM-Newton} (Tamura et al. 2004). 
We also show that the predicted evolution in the Iron
abundance, which should decrease by a factor of 2 at $z \approx 1$
with respect to the local value, is in good agreement 
with the current observational constraints obtained from 
a sample of 49 clusters at $z>0.3$ (see Fig.~\ref{fig:fe_z}).

By using these models to describe the ICM enrichment, we can infer
the relative SN contribution to the amount of elements
present in galaxy clusters as function of the cosmic time. 
At increasing redshift, the products from SNe CC become
dominant, owing to the steep rise of their relative rate
with respect to SNe Ia. 
The transition occurs between $z=0.5$ and
$1.4$, with an enhancement of the 
$\alpha-$elements with respect to Fe (e.g. O/Fe and Si/Fe
ratios increase by a factor of more than 2) and a drastic
decrease of the Ni/Fe ratio (see Fig.~\ref{fig:yfe_z}).
Then, the fractions of metals mass due to SN Ia outputs,
which are locally about 51, 75, 1, 14 and 21 per cent of the total
for Iron, Nickel, Oxygen, Silicon and Sulfur, respectively,
halve at $z \approx 1$ (Fig.~\ref{fig:mia_z}), 
almost independently from the adopted delay time distribution 
functions and models for SN CC explosions.
When the assumed SN rates (number per comoving volume per rest-frame year)
are converted to units of the B-band luminosity (but see caveats in Dahlen
et al. 2004), which is well mapped only in nearby galaxy clusters, 
we show that the expected total $M_{\rm Fe} / L_B$ is still below
the local measurement by a factor between 3 and 5 and that 
SN Ia metal production contributes by $\sim$70 per cent.
We conclude that this well-known (e.g. Arnaud et al. 1992, Renzini et al. 1993)
underestimate of the local $M_{\rm Fe} / L_B$ value cannot be explained
with an increase in the SN rates at higher redshift, according to the 
present models.

The changes by a factor of 2 in the abundance ratios at $z>0.5$ 
arising from the predominance of the enrichment through SNe CC
might be investigated in the near future with X-ray spectroscopically
resolved metal abundance estimates in high-redshift galaxy clusters.
We are attempting this with our {\it Chandra} and {\it XMM-Newton}
data set of high$-z$ objects (Tozzi et al. 2003, Ettori et al. 2004a), with
expected relative uncertainties on the abundance ratios of larger than 
$\sim20$ per cent at the $1 \sigma$ level (see, for example, Fig.~\ref{fig:fe_z}). 
These X-ray satellites offer the best compromise available at present
between field-of-view, effective area, and the spatial and spectral 
resolution required to pursue such a study.
Only with {\it XEUS}\footnote{\tt http://www.rssd.esa.int/index.php?project=XEUS} 
and its sensitivity greater by two orders of magnitude than that 
of {\it XMM-Newton} and a spectral resolution of the order of 10 eV or less,
will it be possible to investigate with significantly higher accuracy the metal
budget of the ICM in high$-z$ systems.

\section*{ACKNOWLEDGEMENTS}  
I thank Stefano Borgani for insightful comments and suggestions, 
Alexis Finoguenov and Paolo Tozzi for useful discussions.
Italo Balestra is thanked for the spectral analysis of the data
presented in Fig.~\ref{fig:fe_z}.
I thank an anonymous referee for helpful remarks relevant to 
improving the presentation of this work.


\begin{thebibliography}{} 

\bibitem{} Anders E., Grevesse N., 1989, Geochimica et Cosmochimica Acta, 53, 
197
\bibitem{} Arnaud M., Rothenflug R., Boulade O., Vigroux L., Vangioni-Flam E., 1992, 
A\&A, 254, 49
\bibitem{} Arnaud M., Pointecouteau E., Pratt G.W., 2005, A\&A, in press (astro-ph/0502210)
\bibitem{} Baumgartner W.H., Loewenstein M., Horner D.J., Mushotzky R.F., 
2005, ApJ, 620, 680
\bibitem{} B\"ohringer H. et al., 2001, A\&A, 365, L181
\bibitem{} Cappellaro E. et al., 2005, A\&A, 430, 83
\bibitem{} Dahlen T. et al., 2004, ApJ, 613, 189
\bibitem[]{} De Grandi S., Ettori S., Longhetti M., Molendi S., 2004, A\&A, 419, 7 
\bibitem[]{} De Grandi S., Molendi S., 2002, in ASP Conf. Proc. 253, 
Chemical Enrichment of Intracluster and Intergalactic Medium, ed. Fusco-Femiano
\& Matteucci (San Francisco), 3
\bibitem{} Dupke R.A., Arnaud K.A., 2001, ApJ, 548, 141
\bibitem[]{} Ettori S., Fabian A.C., Allen S.W., Johnstone R., 2002, MNRAS, 331, 635
\bibitem[]{} Ettori S., Tozzi P., Borgani S., Rosati P., 2004a, A\&A, 417, 13
\bibitem[]{} Ettori S., Borgani S., Moscardini L., Murante G., Tozzi P., Diaferio A., Dolag K.,
Springel V., Tormen G., Tornatore L., 2004b, MNRAS, 354, 111
\bibitem[]{} Finoguenov A., David L.P., Ponman T.J., 2000, ApJ, 544, 188
\bibitem[]{} Finoguenov A., Matsushita K., B\"ohringer H., Ikebe Y., 
Arnaud M., 2002, A\&A, 381, 21
\bibitem[]{} Fukazawa Y., Makishima K., Tamura T., Ezawa H., Xu H., Ikebe Y., 
Kikuchi K., Ohashi T., 1998, PASJ, 50, 187
\bibitem[]{} Gal-Yam A., Maoz D., 2004, MNRAS, 347, 942
\bibitem[]{} Gastaldello F., Molendi S., 2002, ApJ, 572, 160
\bibitem[]{} Gastaldello F., Molendi S., 2004, ApJ, 600, 670
\bibitem[]{} Gibson B.K., Loewenstein M., Mushotzky R.F., 1997, MNRAS, 290, 623
\bibitem[]{} Grevesse N., Sauval A.J., Space Science Rev., 85, 161
\bibitem[]{} Hashimoto Y., Barcons X., B\"ohringer H., Fabian A.C., Hasinger G., 
Mainieri V., Brunner H., 2004, A\&A, 417, 819
\bibitem[]{} Lin Y.-T., Mohr J.J., Stanford S.A., 2003, ApJ, 591, 749
\bibitem[]{} Maoz D., Gal-Yam A., 2004, MNRAS, 347, 951
\bibitem[]{} Matteucci F., Vettolani G., 1988, A\&A, 202, 21
\bibitem[]{} Mitchell R.J., Culhane J.L., Davison P.J.N., Ives J.C., 1976, 
MNRAS, 175, 29P
\bibitem[]{} Mushotzky R.F., Loewenstein M., Arnaud K.A., Tamura T., Fukazawa Y.,
Matsushita K., Kikuchi K., Hatsukade I., 1996, ApJ, 466, 686
\bibitem[]{} Nakamura T., Umeda H., Nomoto K., Thielemann F.-K., Burrows A., 
1999, ApJ, 517, 193
\bibitem[]{} Nomoto K., Iwamoto K., Nakasato N., Thielemann F.-K.,
Brachwitz F., Tsujimoto T., Kubo Y., Kishimoto N., 1997, Nucl. Phys. A,
621, 467 (astro-ph/9706025)
\bibitem[]{} Portinari L., Moretti A., Chiosi C., Sommer-Larsen J., 2004, ApJ,
604, 579
\bibitem[]{} Renzini A., Ciotti L., D'Ercole A., Pellegrini S., 1993, ApJ, 419, 52 
\bibitem[]{} Renzini A., 2003, (astro-ph/0307146)
\bibitem[]{} Rosati P. et al., 2004, AJ, 127, 230
\bibitem[]{} Salpeter E.E., 1955, ApJ, 121, 161
\bibitem[]{} Sanders J.S., Fabian A.C., Allen S.W., Schmidt R.W., 2004, MNRAS, 
349, 952
\bibitem[]{} Serlemitsos P.J., Smith B.W., Boldt E.A., Holt S.S., Swank J.H., 
1977, ApJ, 211, L63
\bibitem[]{} Strolger L.G. et al., 2004, ApJ, 613, 200
\bibitem[]{} Tamura T., Kaastra J.S., den Herder J.W.A., Bleeker J.A.M., Peterson J.R.,
2004, A\&A, 420, 135
\bibitem{} Tozzi P., Rosati P., Ettori S., Borgani S., Mainieri V., Norman C., 
2003, ApJ, 593, 705
\bibitem{} Woosley S.E., Weaver T.A., 1995, ApJS, 101, 181

\end{thebibliography}
\end{document}